\documentclass[11pt]{article}
\pdfoutput=1 
\usepackage{jheppub} 
\usepackage[T1]{fontenc} 

\usepackage{graphicx}
\usepackage{float}
\usepackage{hyperref}

\title{Leptonic number violation signature arising from a $Z'$ model 
with non-diagonal leptonic couplings}

\author{J. M. Hern\'andez-L\'opez, }
\author{T. A. Valencia-P\'erez}  
\affiliation{Facultad de Ciencias F\'isico Matem\'aticas, B. Universidad 
Aut\'onoma de Puebla, Apdo. Postal 1364, 72000, Puebla, M\'exico.}
\emailAdd{javierh@fcfm.buap.mx}
\emailAdd{214702210@alumnos.fcfm.buap.mx}

\abstract{Within the framework of a $Z'$ model with non-universal leptonic 
couplings, we analyze possible signatures of leptonic number violation 
effects at LHC.  Results are described for leptonic energy distributions, both 
from its specific signature and events number, that could allow us to 
observe this class of models, under reasonable conditions at LHC.}

\keywords{$Z'$, LFV}

\begin{document}
\maketitle
\flushbottom

\section{Introduction}
\label{sec:level1} 

Although the Standard Model (SM) has been succesful in describing the physics at 
the Electroweak (EW) scale, and the particle content of the EW interactions can 
be considered completed with the recent Higgs boson discovery 
\cite{Aad:2012tfa,CMS:2012xaa}, it is considered as the low energy limit of a 
more fundamental theory. So New Physics (NP) should exist at energy scales of a 
few TeV. Among the many models of physics beyond the SM, there exist a class of 
models that predict the presence of a structure like 
$SU(2)_1\times SU(2)_2\times U(1)_X$ at the scale of a few TeVs
\cite{Mohapatra:1974hk}\cite{He:000he,Yue:2002ja,Hsieh:2010zr,Langacker:2008yv}. 
Such a gauge symmetry may have its origin from various grand unified theories, 
string-inspired models, and even dynamical symmetry breaking models, and little 
Higgs models. See, for example \cite{Langacker:2008yv} and references therein.
In these class of models, there exist new gauge bosons. In particular, new neutral
gauge bosons that are going to be searched in all present and future colliders
\cite{Langacker:2008yv,Diener:2009vq}. 

Usually, flavor-conserving $Z'$ bosons have been extensively studied
\cite{Langacker:2008yv,Hsieh:2010zr}, while less
attention has been given to the more general case of $Z'$ bosons as a primal
source of Lepton Flavour Violation (LFV) processes.  One of the features of such
kind of models is the presence of a heavy $Z'$ gauge bosons with mass in the TeV scale, well within the reach of LHC \cite{Langacker:2000ju,Zhang:2012jm}. 
Also there have been some work on a global EW fit that could prefers a 
$Z'$ which do not couple to the first leptonic and second quark generation \cite{Zhang:2012jm}, leading to family dependent fermionic couplings for a new $Z'$.
That opens an opportunity windows for the search of $Z'$ contributions in the
case of non-diagonal fermionic couplings. Following that work,
the aim of our paper is to study leptonic number violation effects at LHC 
through the processes  $pp \to Z'\longrightarrow l_{i}\bar{l}_{j}$ and  $pp \to Z' \longrightarrow l_{i}\bar{l}_{j}h$.


The structure of the paper is the following: in section 2, we are going over the 
details of the model and the processes under consideration. We then describe
the results for differential cross section distributions accordingly. Our conclusions are given in the final section.

\section{Lepton number violation processes}
\label{sec:level2}

In this paper we would focus in the feasibility of direct detection 
of a heavy Z' gauge boson through the distintive signature of lepton number
violation plus $h$ boson. We are, first, assuming that a $SU\left(2\right)_{1}\times SU\left(2\right)_{2}\times U\left(1\right)_{X}$ gauge structure represents the correct physics at the TeV scale \cite{Hsieh:2010zr}. Based on that
assumption, we allow for the possibility that the fermion-$Z'$ couplings be 
non-universal, as in the model from Langacker-Plumacker~\cite{Langacker:2000ju}. 
This lead us to LFV effects, which we are going to focus; aswell as FCNC 
contributions which are not treated in the present paper. In order to scan the
parameter space in a more efficient way, we parametrise the 
non-universal fermionic-$Z'$ coupling in an effective lagrangian way (see for
example \cite{FloresTlalpa:2001sp}). Moreover, in order to restrict the number
of unknown parameters, only the leptonic sector it is considered to be non-
diagonal, while the quark sector couplings are considered to behave as
SM-like.

The Feynman diagrams for the considered processess,

\centerline{ 
1. $pp\longrightarrow Z^{\prime}\longrightarrow l_{i}\bar{l}_{j}X$, \hspace{1.5cm}
2. $pp\rightarrow Z\rightarrow Z^{\prime}h\rightarrow l_{i}\bar{l}_{j} h X$
}

\noindent are shown in the following figures.

\begin{figure}[H]
\centering
\parbox{6cm}{%
\includegraphics[scale=0.13]{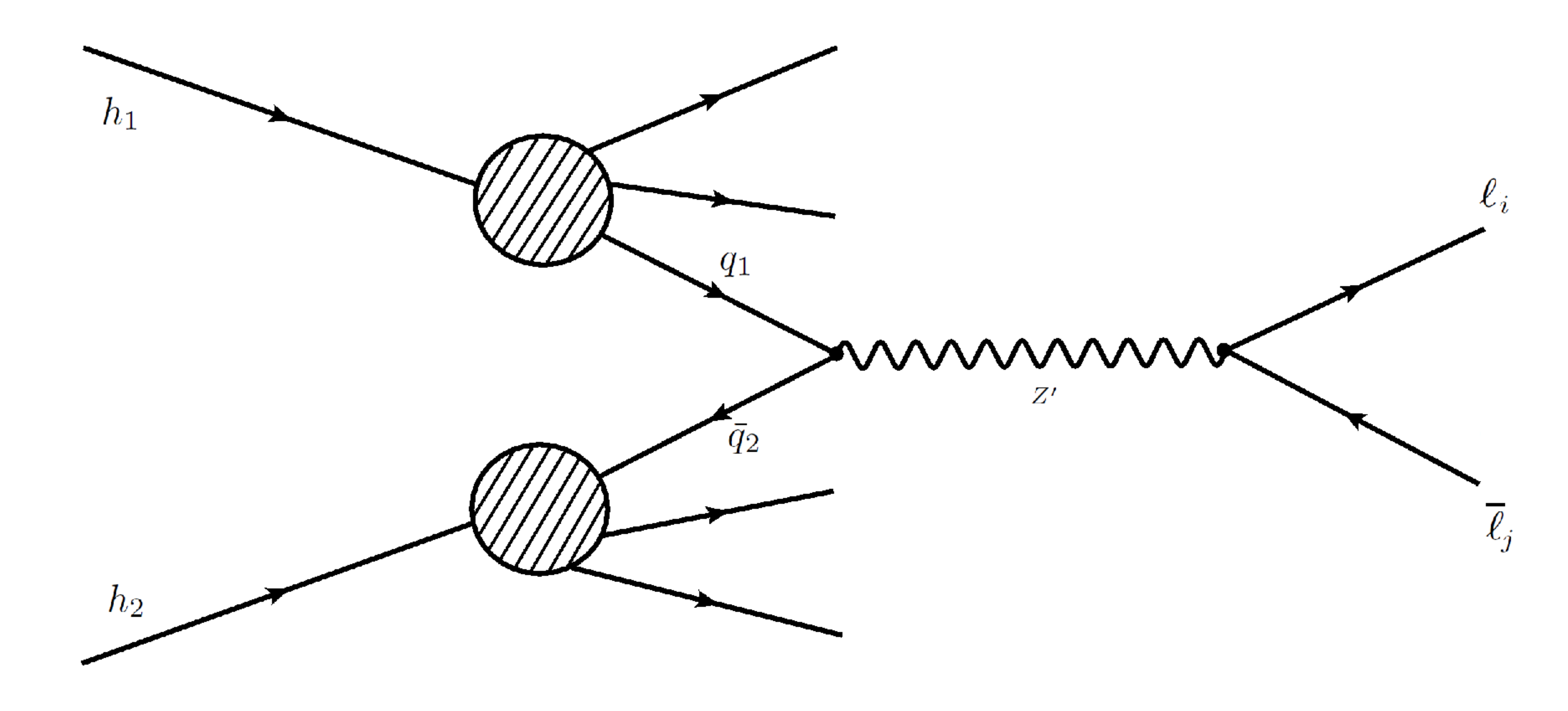}
\caption{\label{fig:proceso1_hadrones}Feynman diagrams for the process 1.}}
\qquad
\begin{minipage}{6cm}%
\includegraphics[scale=0.09]{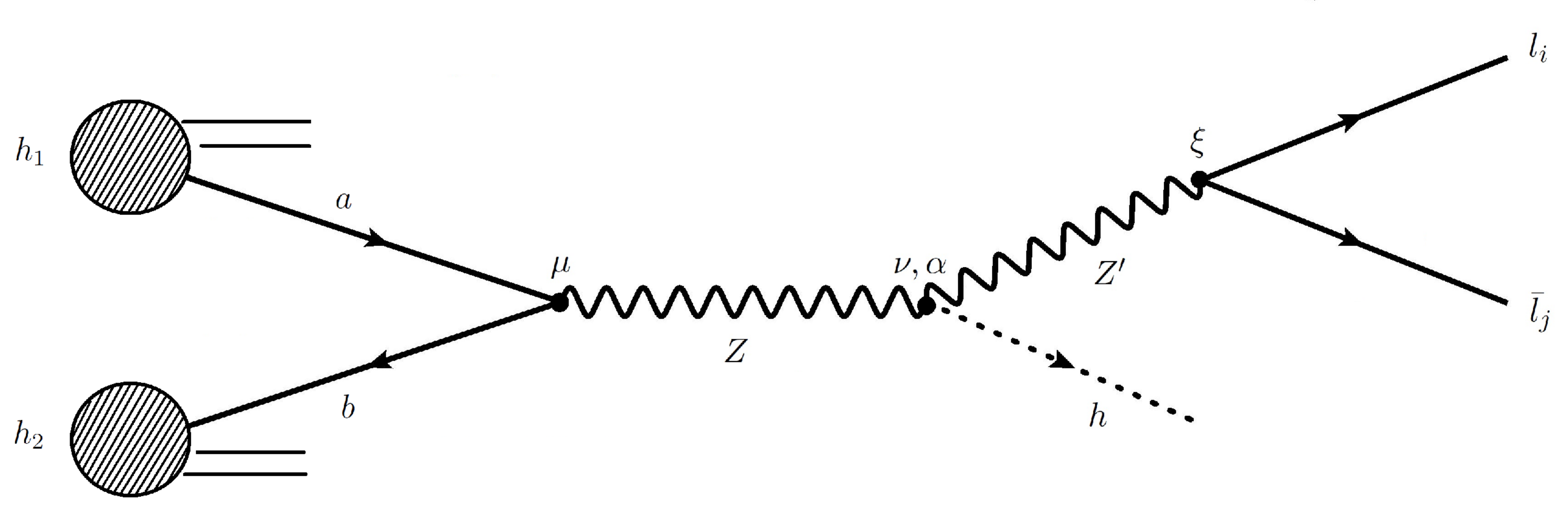}
\caption{\label{fig:proceso2_hadrones}Feynman diagrams for the process 2.}
\end{minipage}
\end{figure}

In the following 
we will designate as Process 1 to $pp \to Z'\longrightarrow l_{i}\bar{l}_{j}X$, while Process 2 will be used for $pp \to Z' \longrightarrow l_{i}\bar{l}_{j}hX$.
The leptonic energy distributions were done, as remarked already, by using an 
effective approach for the vertex  $Z^{\prime}l_{i}\bar{l}_{j}$, parametrised as 
$C_{V}\left(g_{Vij}^{f}-g_{Aij}^{f}\gamma_{5}\right)\gamma^{\nu}$.
While the small vertex $ZZ^{\prime}h$ is given by the Lorentz structure 
$C_{ZZ^{\prime}h}g^{\nu\alpha}$. Given the 
non-universality of the fermionic couplings of the $Z'$, the $g_{Vij}^{f}$ and 
$g_{Aij}^{f}$ are matrices, generally with all its elements not null.
For the sake of simplifying the number of free parameters to be considered, the 
quark-$Z^{\prime}$ vertex was set both as diagonal in the flavour space and with 
its coupling equal as in the SM. The coefficient of the fermionic-$Z'$ coupling, 
$C_V= .8 C_{SM}$, for definitiveness, and the coupling  
$C_{ZZ^{\prime}h} = 0.01-.001$, given the restrictions that arise usually from 
the $Z-Z'$ mixing.

In order to consider the future behaviour of LHC, we have considered both cases 
of $\sqrt{s}=8$, and 14 TeV center of mass (CM) energy. 
We analyse some interesting differential distributions for the final particles, 
as a way to disentangle the most interesting signatures that allow us a direct 
path to found the $Z'$ signal at LHC. Given the different possibilities for the
detection of the Higgs boson, we use the $\gamma \gamma$ decay case, and consider
the Process 2 as $pp \to Z' \longrightarrow l_{i}\bar{l}_{j} \gamma \gamma X$.

\section{Results for LFV Processes }

This section will show the obtained results for several differential
distributions of the corresponding cross sections for Processes 1 and 2.
We will focus in the signal at the resonance, since we have found that give us 
the maximum signal for the processes under study. 

For the Process 1, namely $pp \to Z' \to l_{i} \bar
{l}_j$, a Drell-Yan-like process, we calculate both the differential cross 
section  with respect to the reduced partonic energy, $\tau$, aswell as the 
expected number of events for the $l_{i} \bar{l}_j$ invariant mass.

Figure (3) shows the differential cross section as a function of the reduced
partonic energy, the ratio of the quark CM energy vs $pp$ CM energy.
Because LHC has a strong limit on SM-like fermionic coupling for sequential 
models of $Z'$, we use in our analysis masses of 2, 2.5 and 3 TeV for the $Z'$.
The figure at the left corresponds to the case $\sqrt{s}=8$ TeV, while the 
right figure corresponds to 
$\sqrt{s}=16$ TeV. In the plots involving $\tau$, a cut was made at $10^-3$.
The library LHAPDF 
v 5.8.8 was used to define the corresponding PDFs, with CTEQ 6 \cite{Whalley:2005nh}.

\begin{figure}[h]
\centering
\parbox{6.5cm}{%
\includegraphics[angle=-90,scale=0.28]{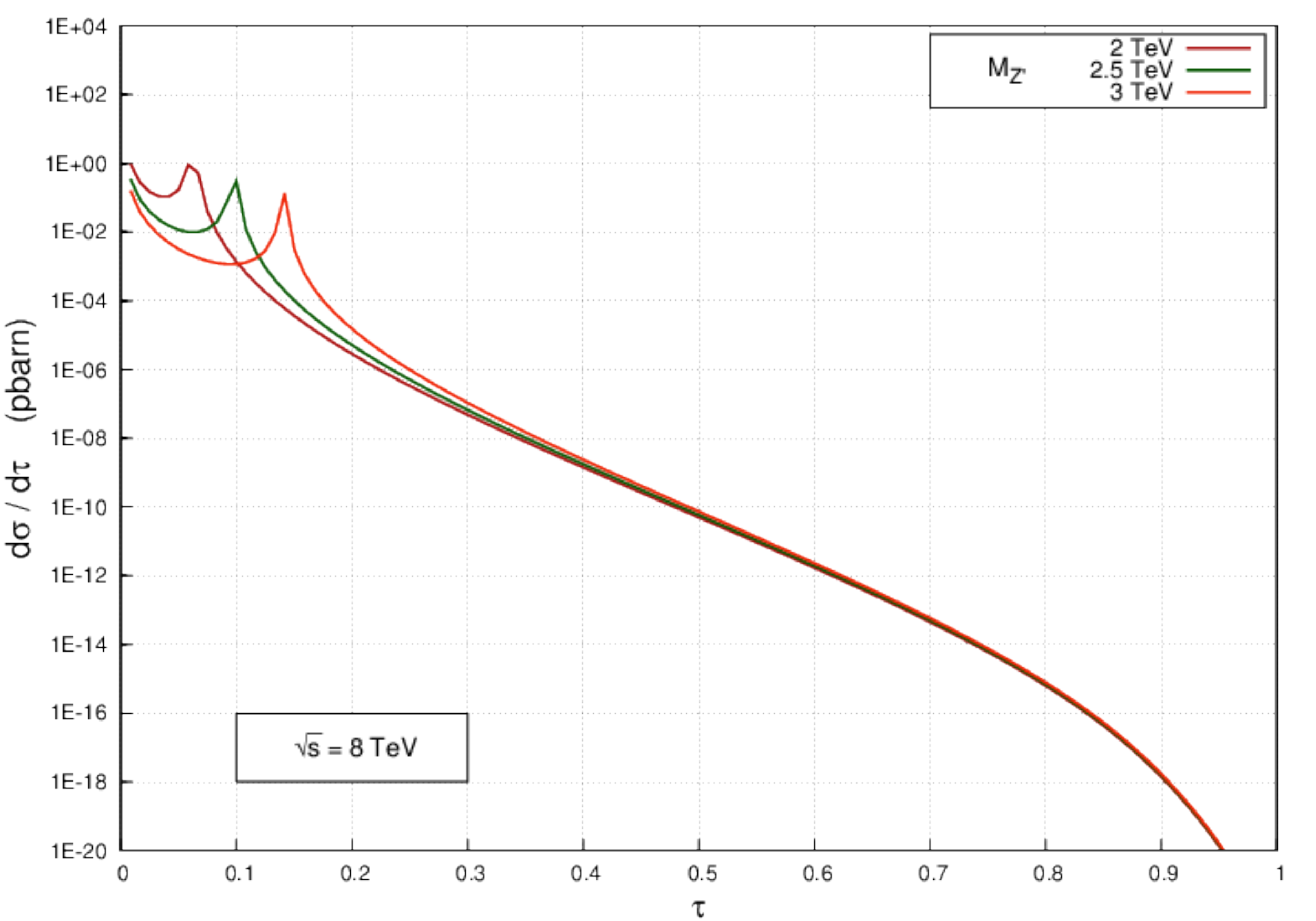}}
\qquad
\begin{minipage}{6.5cm}%
\includegraphics[angle=-90,scale=0.28]{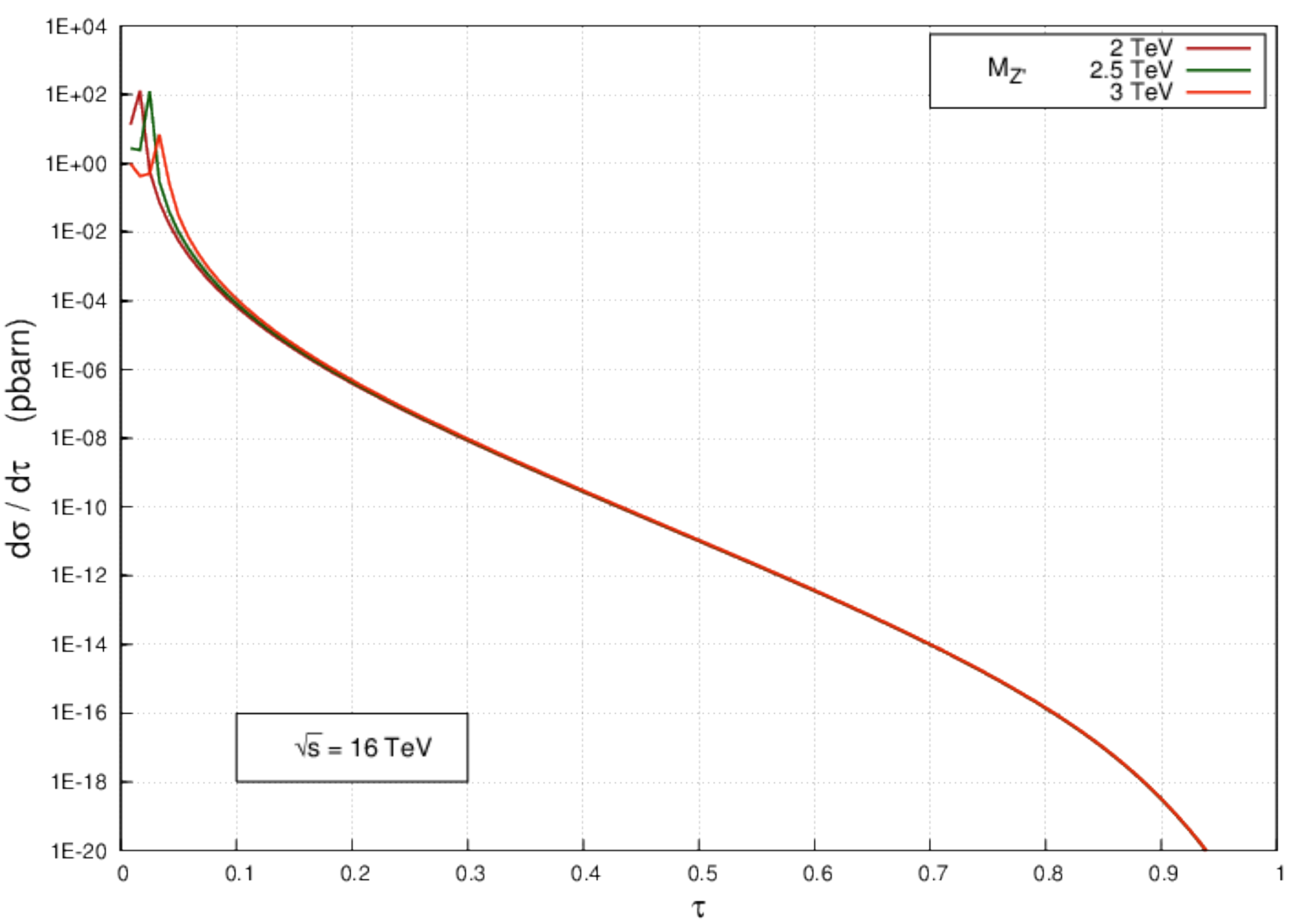}
\end{minipage}
\caption{\label{fig:Energia-Reducida-1}Process 1: Differential cross section vs. reduced partonic energy, at $\sqrt{s} = 8$ (left), and 16 TeV (right). We present
the plots for $M_{Z'} = 2, 2.5,$ and $3$ TeV.}
\end{figure}

We can appreciate the signal is only relevant near the resonance $Z'$ peak, and
that happens at low values for $\tau$, both for $\sqrt{s}=8, 16$ TeV.

Figure~(4) shows the same differential distribution, 
but in this case we constrast the effects at $\sqrt{s}=8$, and at 16 TeV, for
a $Z'$ of mass 2.5 (left), and 3 (right) TeV, correspondingly. 
So we can expect a $10^{-2}$ factor for the 16 TeV case, compared with the 8 TeV
case. 

\begin{figure}[h]
\centering
\parbox{6.5cm}{%
\includegraphics[angle=-90,scale=0.28]{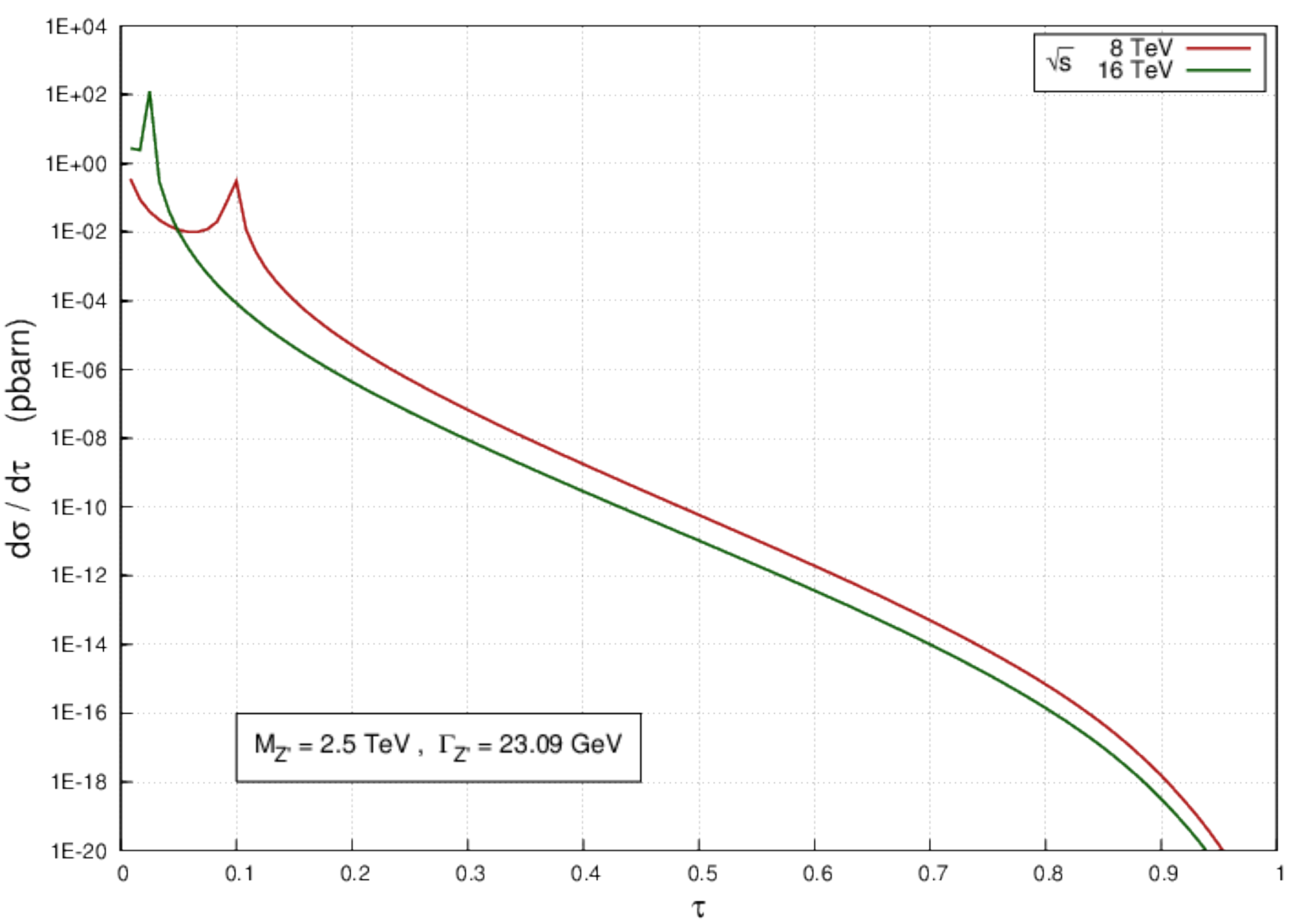}}
\qquad
\begin{minipage}{6.5cm}%
\includegraphics[angle=-90,scale=0.28]{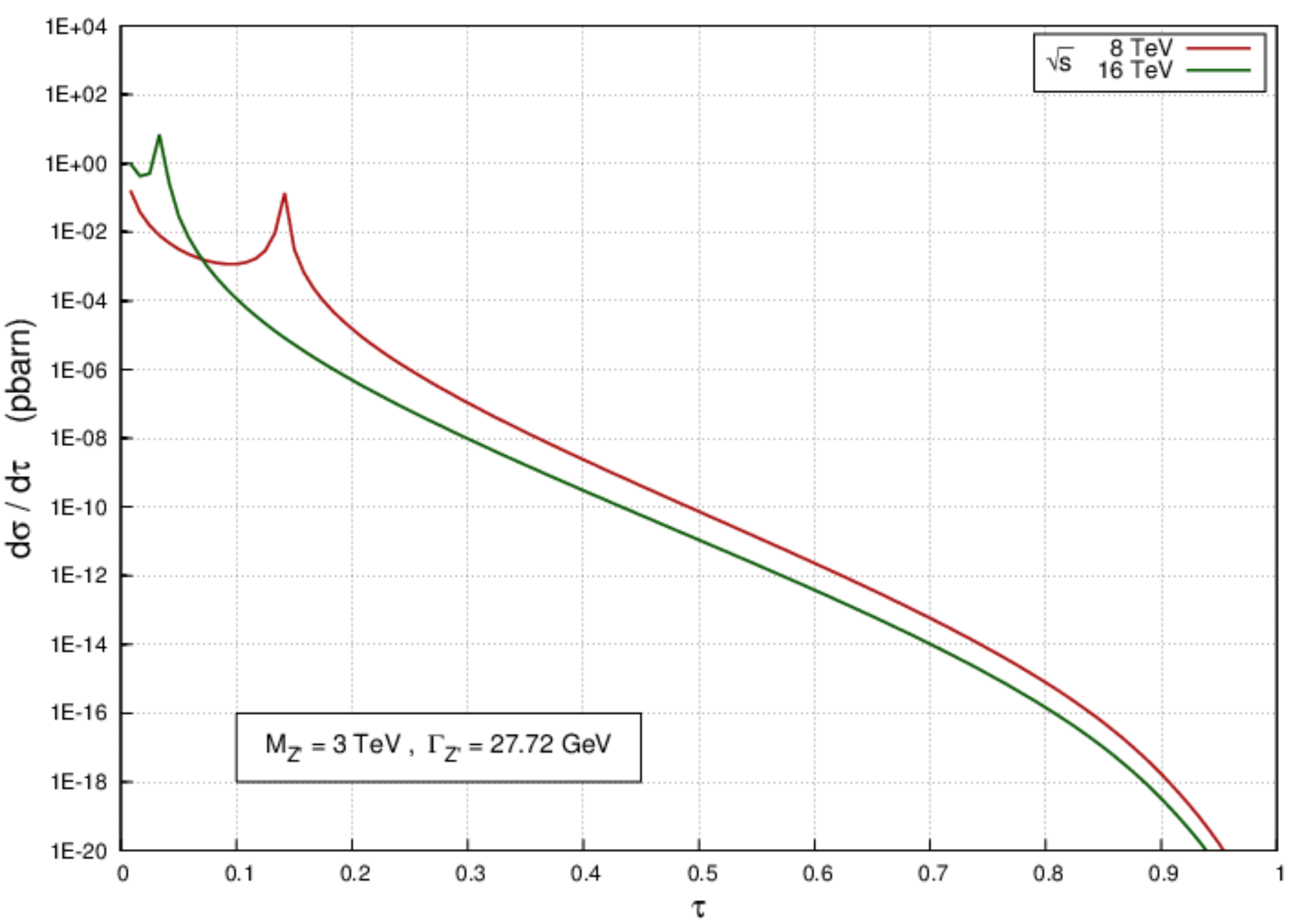}
\end{minipage}
\caption{\label{fig:Energia-Reducida-3}Process 1: Differential cross section vs. 
reduced partonic energy, for $M_{Z'}=2.5$ (left), and 3 (right) TeV. The plots
compare the cases of $\sqrt{s} = 8$, and 16 TeV.}
\end{figure}

In order to quantify if this kind of models can give us an observable number of events, we assume reference values from LHC. 
For the number of events, we take into consideration the electron and muon 
detection eficency, taking nominally the same numbers for both ATLAS and CMS 
\cite{Atlas-CMS:2013,Laurent:2013lt}. The reported luminosity for the 8 
TeV run is 23.269 fb${}^{-1}$ \cite{Luminosity:2012}.  
We have used a luminosity of 60-100 $fb^{-1}$ for a  14 TeV LHC CM energy because 
the experimental expectatives at that energies \cite{Luminosidad:LHC14}. Also in Ref.~\cite{Mansour:2012ek} have been pointed that the possibilities for Z' 
discovering at a future LHC relies in that levels of luminosity.
\begin{figure}[hb]
\centering
\parbox{4cm}{%
\includegraphics[angle=0,scale=0.24]{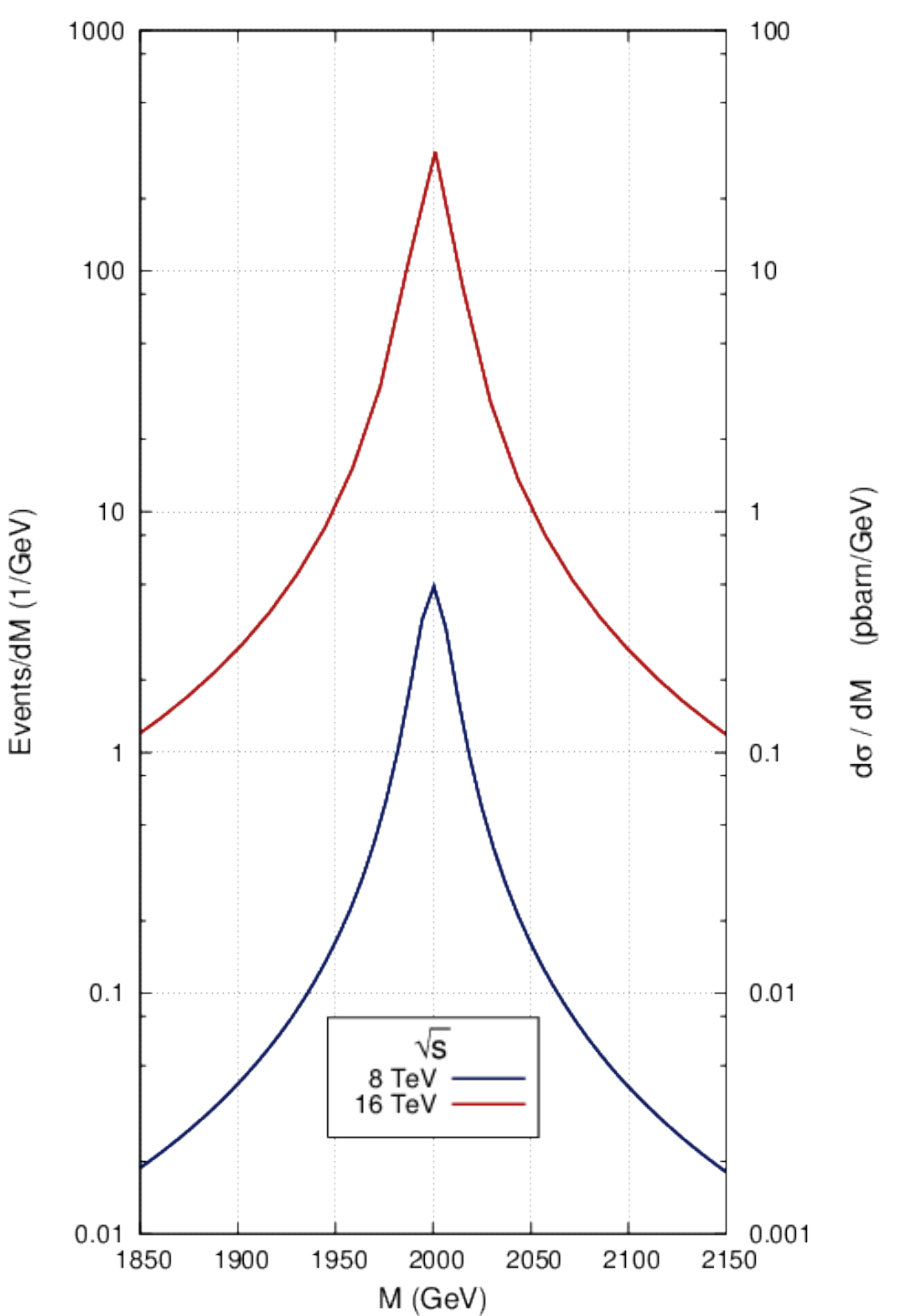}}%
\qquad
\begin{minipage}{4cm}%
\includegraphics[angle=0,scale=0.24]{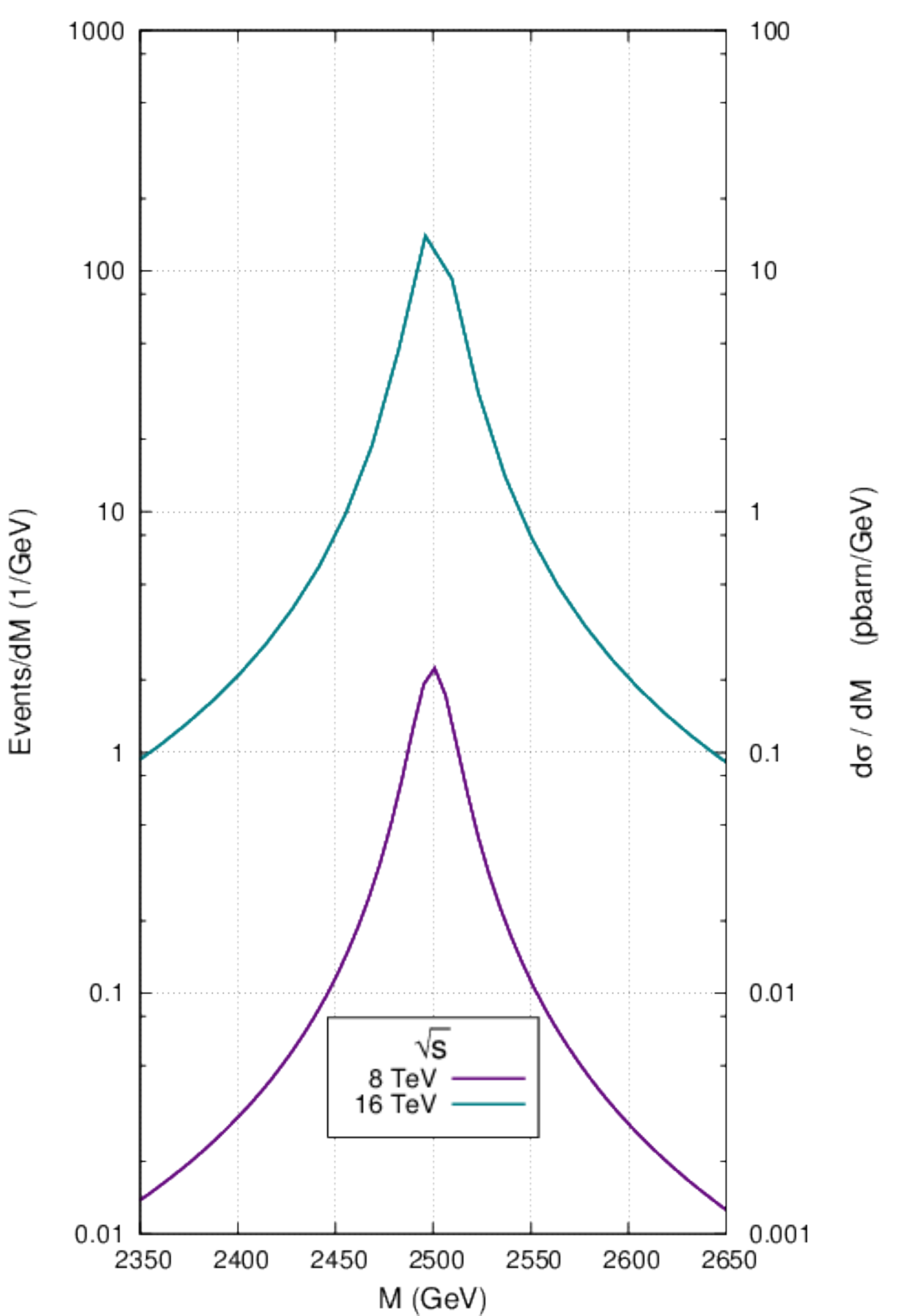}
\end{minipage}
\qquad
\begin{minipage}{4cm}%
\includegraphics[angle=0,scale=0.24]{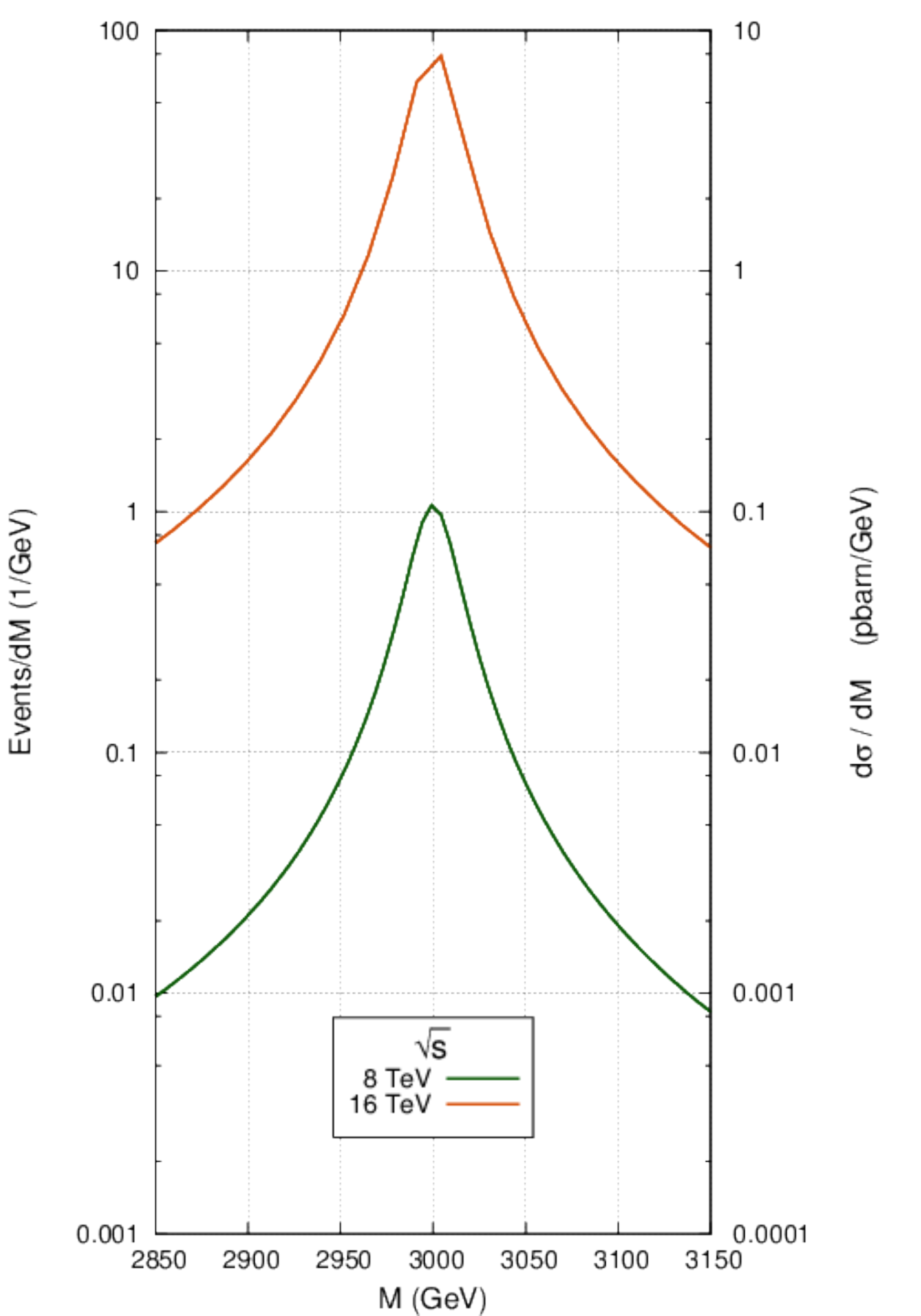}
\end{minipage}
\caption{\label{fig:Eventos-2}Process 1: Differential distributions for the 
number of events (left scale) vs. leptonic invariant mass, and differential cross 
section (right scale) vs. leptonic invariant mass for the cases of $M_{Z'}$=2 
(left), 2.5 (center), and 3 (right) TeV. We show the dependence for 
$\sqrt{s} = 8$, and 16 TeV. The conditions for the luminosity are specified in the text.}
\end{figure}

\begin{figure}[h]
\centering
\parbox{3.5cm}{ 
\includegraphics[angle=0,scale=0.23]{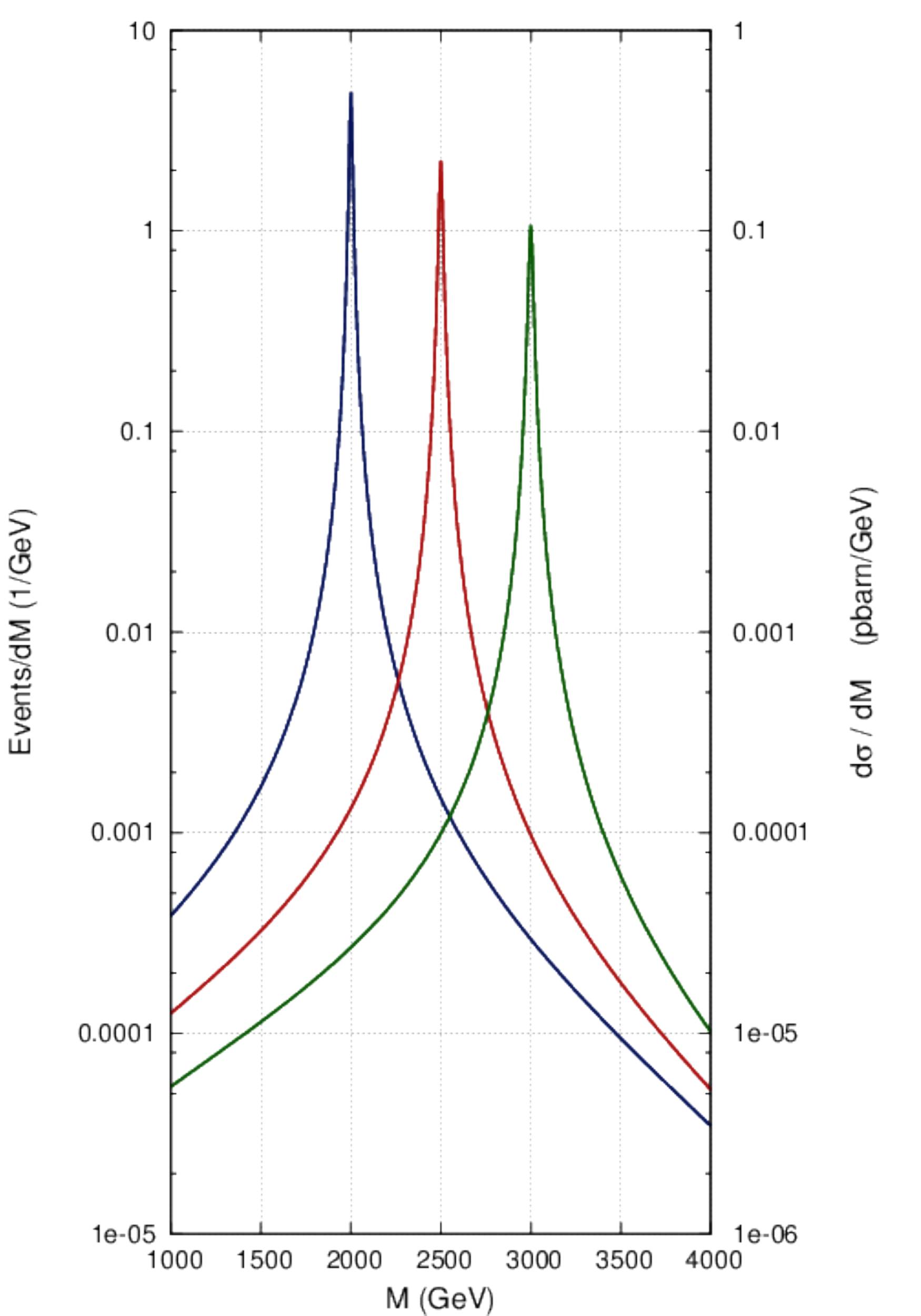}}
\qquad
\begin{minipage}{3.5cm}%
\includegraphics[angle=0,scale=0.23]{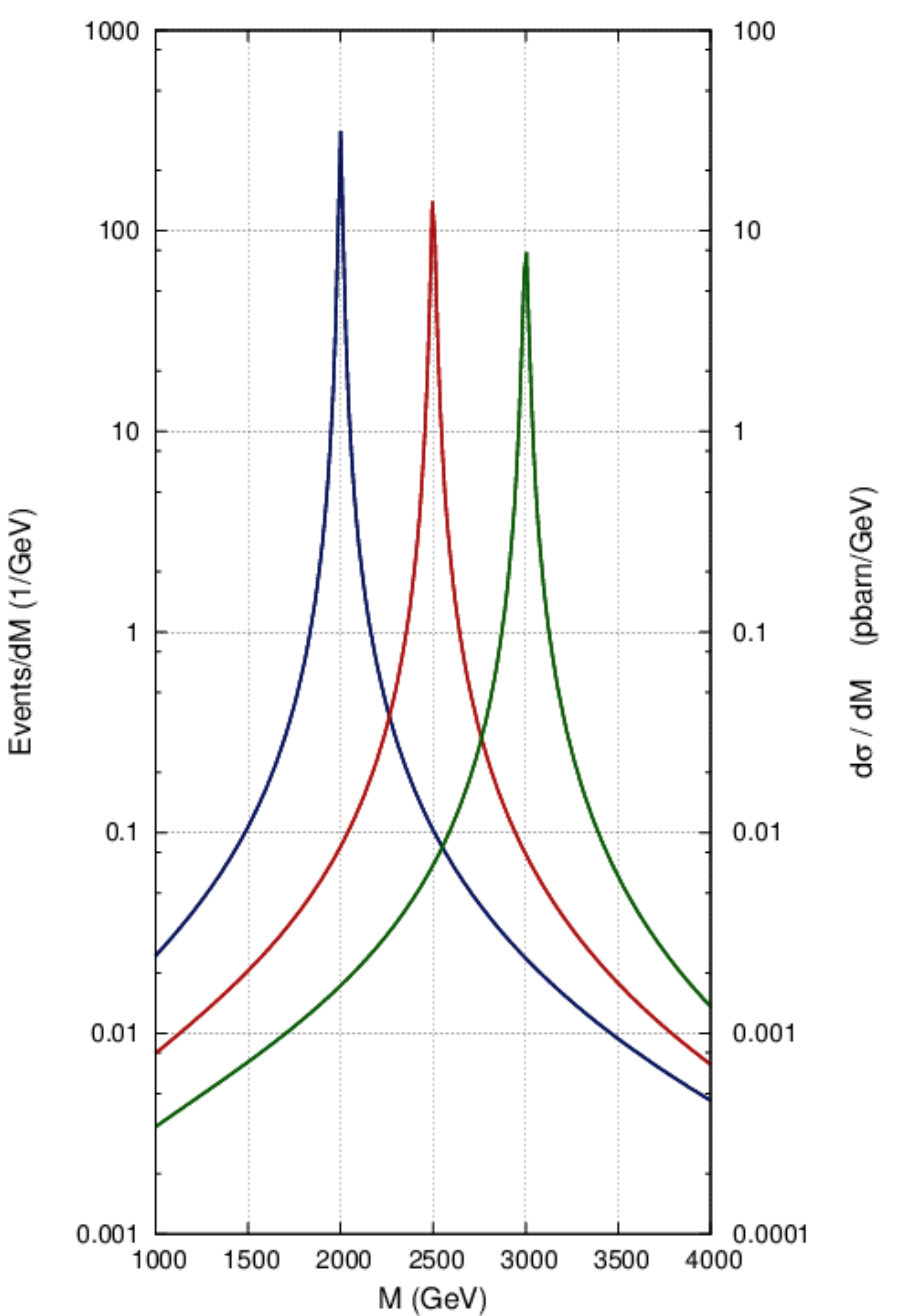}
\end{minipage}
\caption{\label{fig:Eventos-4}Process 1: Differential distribution for the number 
of events (left scale) vs. leptonic invariant mass,  and differential cross 
section (right scale) vs. leptonic invariant mass for the cases of $\sqrt{s} = 8$ 
(left), and 16 (right) TeV. for each graph we conmpare the cases of $M_{Z'}$=2, 
2.5, and 3 TeV. The conditions for the luminosity are specified in the text.}
\end{figure}

The expected number of events for the cases of $M_{Z'} =
2$ (left), 2.5 (center), and 3 (right) TeV is shown in Fig.~(5). The  
LHC CM energy assumed were $8$, and 16 TeV. We take the leptonic invariant mass 
as the  variable, $e$ and $\mu$ are considered for the final leptonic states.

Meanwhile, the figure(6) presents the same distribution as figure~(5), but in
this case we compare the values for the different $Z'$ masses that we have used,
for each of the cases $\sqrt{s} = 8$ (left), and 16 (right) TeV. We can see
that only around the resonance, it is the cross section above $10^{-2}$ $fb^{-1}$
which is considered a measurable signal. Further studies are needed for off-
resonance signals.

The results shown in figs.~(3-6) were obtained with a particular set of values
for the several unknown parameters. And even when in the literature there exist
some boundings for some of them \cite{Langacker:2000ju,Jezo:2012rm}, as we stated
above, we prefer an effective lagrangian approach. Thenceforth, we are allowed
to let run these parameters. 

\begin{figure}[h]
\begin{center}
\includegraphics[angle=0,scale=0.245]{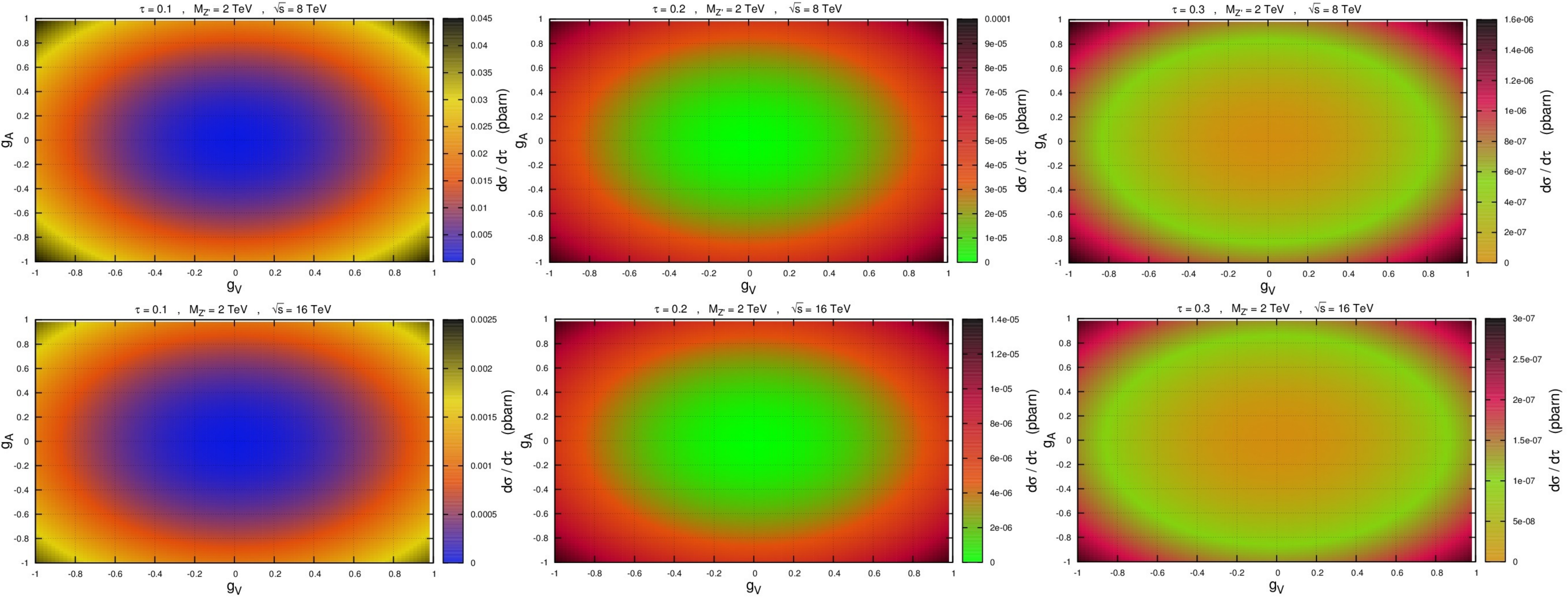}
\caption{\label{fig:gVgA-1}Process 1: Differential cross section as a function of
$\tau$, with the variation on $g_V$ and $g_A$, for the $e-\mu$ decay mode. 
We take $M_{Z'}$=2 TeV for $\tau$ = 0.1 (left), 0.2 (center), and 0.3 (right); 
the first row correspond to $\sqrt{s} = 8$, while the second row for 16 TeV.}
\end{center}
\end{figure}

Figures~(7-9) detail the dependence
on the non-diagonal fermionic couplings, $g_V$ and $g_A$, for the $e-\mu$ decay 
mode of $Z'$. These plots correspond to $M_{Z'} = 2$ (fig.~(7)), 2.5 (fig.~(8))
, and 3 (fig.~(9)) TeV. We plotted the dependence of the differential distribution
for the cross section with respect to $\tau$, but fixing it at .1 (left), .2
(center), and .3 (right) in each of the figures. The fixing was made taking into
account the results shown in Figs.~(3-4). The graphs display the 
corresponding dependence in the $g_V, g_A$ parameter space, for each one of 
the signaled cases. We also present the differences between $\sqrt{s} = 8$
(first line) and 16 (second line) TeV, in each of figures~(7-9).
\begin{figure}[h]
\begin{center}
\includegraphics[angle=0,scale=0.245]{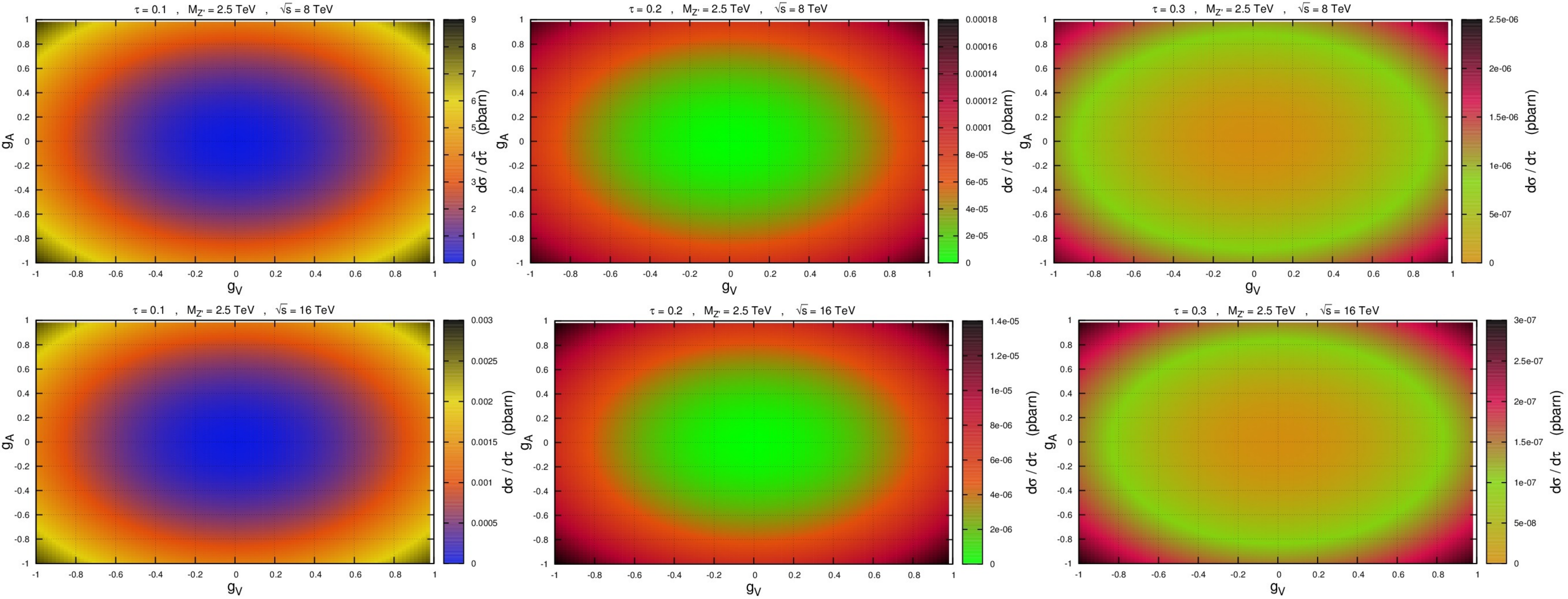}
\end{center}
\caption{\label{fig:gVgA-2}Process 1. Differential cross section with respect to
$\tau$ in function of $g_V$ and $g_A$, for the $e-\mu$ decay mode. 
We take $M_{Z'}$=2.5 TeV for $\tau$ = 0.1 (left), 0.2 (center), and 0.3 (right); 
the first row correspond to $\sqrt{s} = 8$, while the second row for 16 TeV.}
\end{figure}

\begin{figure}[h]
\begin{center}
\includegraphics[angle=0,scale=0.245]{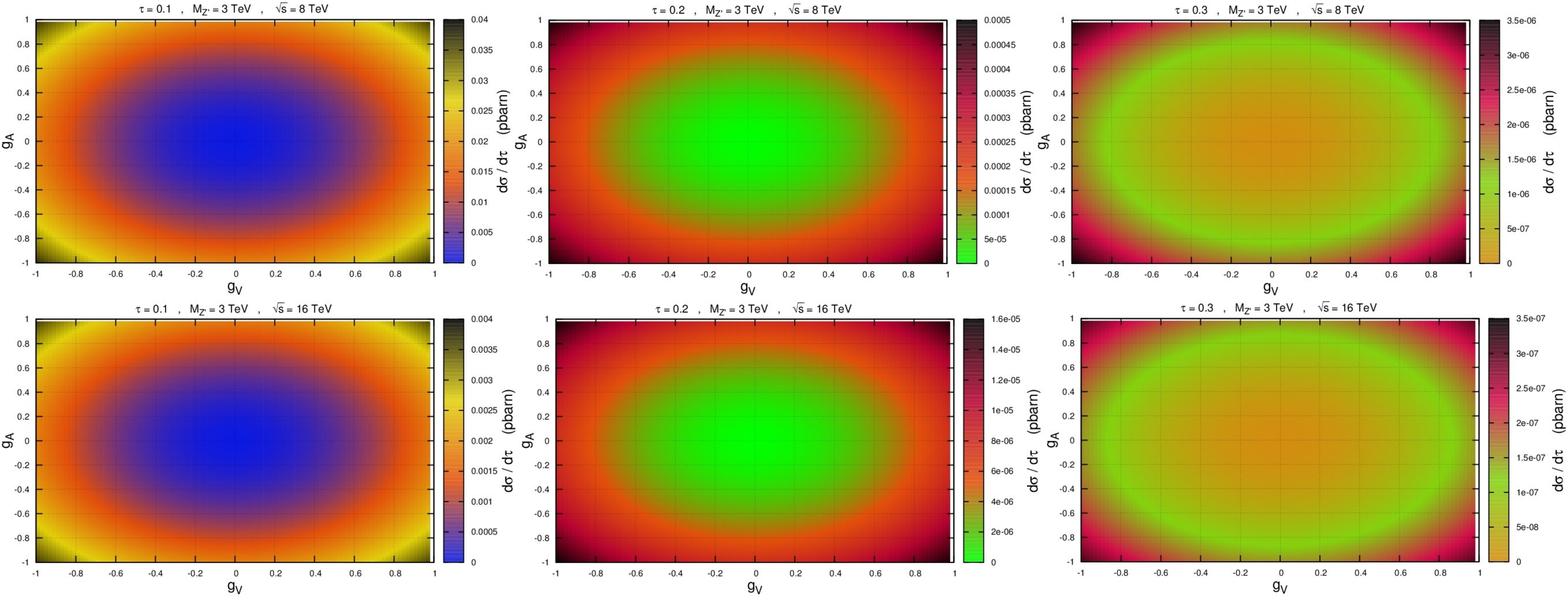}
\end{center}
\caption{\label{fig:gVgA-3}Process 1. Differential cross section vs.  
$\tau$ in function of $g_V$ and $g_A$, for the $e-\mu$ decay mode of the $Z'$. 
We take $M_{Z'}$=3 TeV for $\tau$ = 0.1 (left), 0.2 (center), and 0.3 (right); 
the first row correspond to $\sqrt{s} = 8$, while the second row for 16 TeV.}
\end{figure}

For the second process, which we called Process 2, $pp \to Z \to Z'h \to l_{i}\bar{l}_jhX$, we repeat the procedure for the first process, with the corresponding
changes. We directly introduce the branching ratio of the Higgs boson into 
$\gamma \gamma$, as coming from ATLAS and CMS limits 
\cite{Aad:2012tfa,CMS:2012xaa}. And even when the signal is rather small still
is at the measureable level as at the peak as can be seen in fig.~(10). These 
plots show the 
differential distribution with respect to $\tau$, the reduced partonic energy, 
and $x_{l_i}$, the usually defined $i$-lepton scaled energy ($x_{l_i} = 
E_{l_i}/2\sqrt{s})$. In our case we take $i=e$
\begin{figure}[ht]
\centering
\parbox{4.5cm}{%
\includegraphics[scale=0.22]{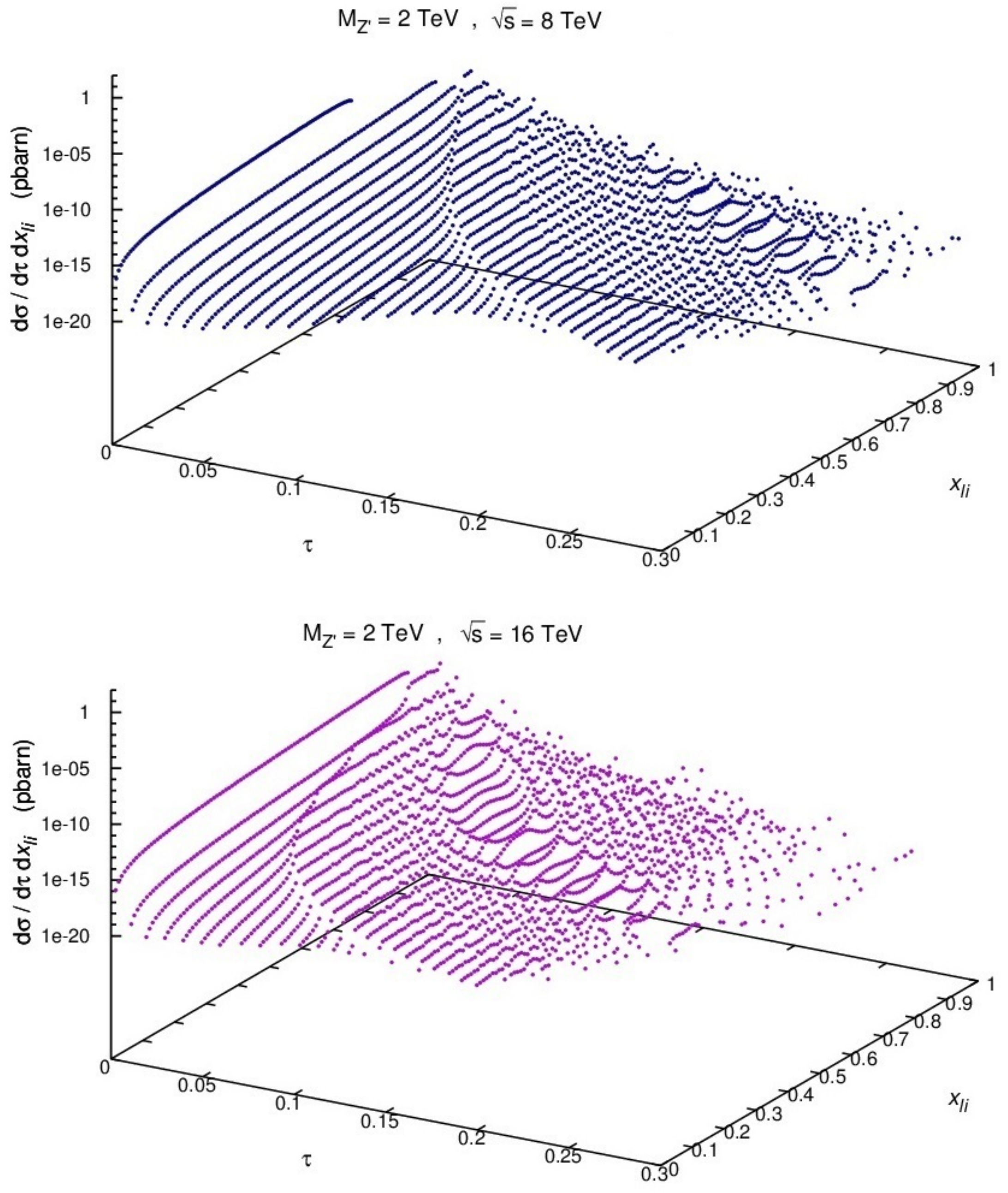}}
\qquad
\begin{minipage}{4.5cm}%
\includegraphics[scale=0.22]{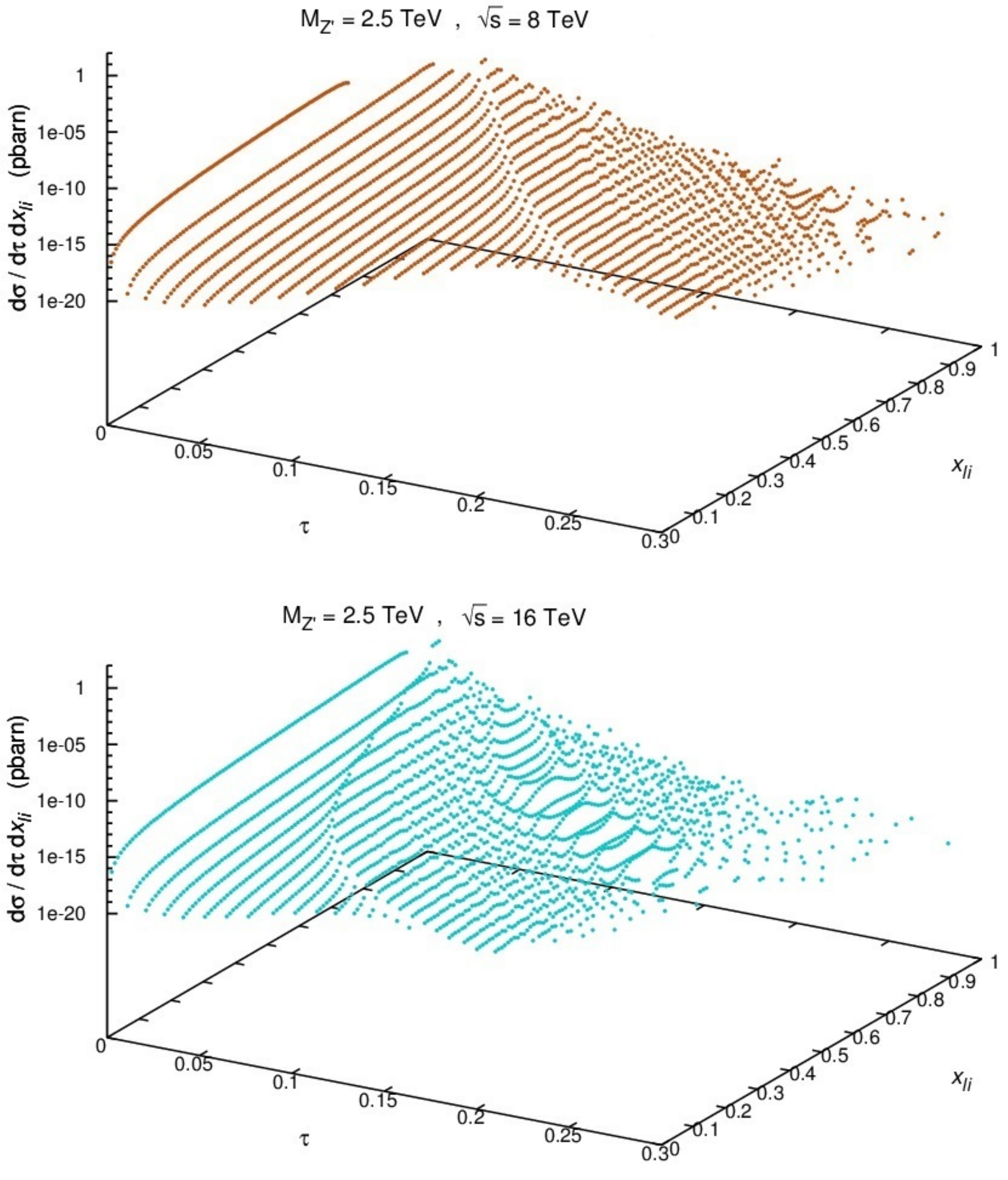} 
\end{minipage}
\qquad
\begin{minipage}{4.5cm}%
\includegraphics[scale=0.22]{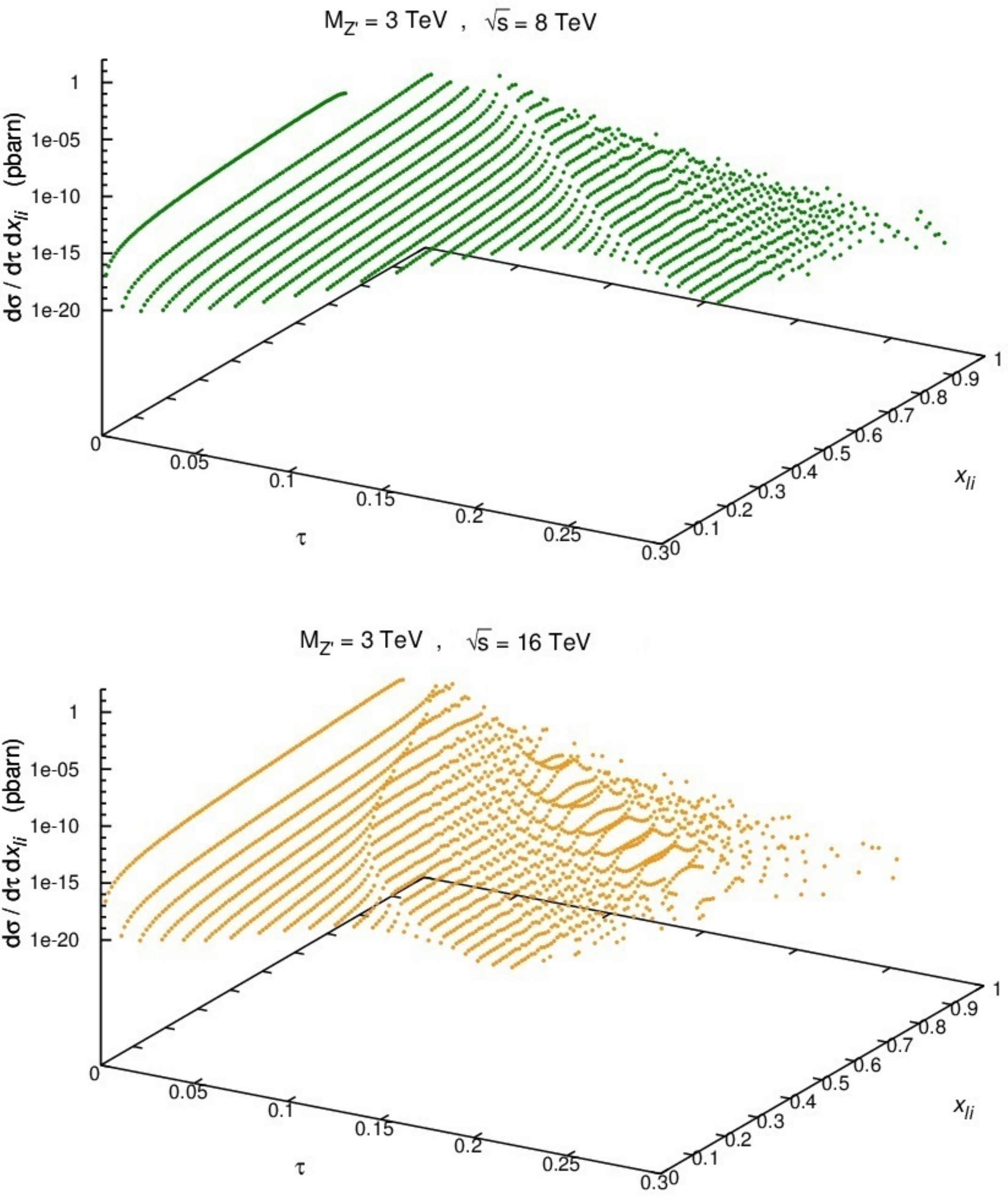} 
\end{minipage}
\caption{\label{fig:Energia-Reducida-6}Process 2: Differential cross section in 
dependence to $\tau$ and electron reduced energy, for $M_{Z'}=2$ (left), and 2.5 
(center), and 3 (right) TeV. The first row of graphs correspond to $\sqrt{s} = 8$
while the second row correspond to the 16 TeV case.}
\end{figure}

\begin{figure}[!Ht]
\centering
\includegraphics[angle=0,scale=0.245]{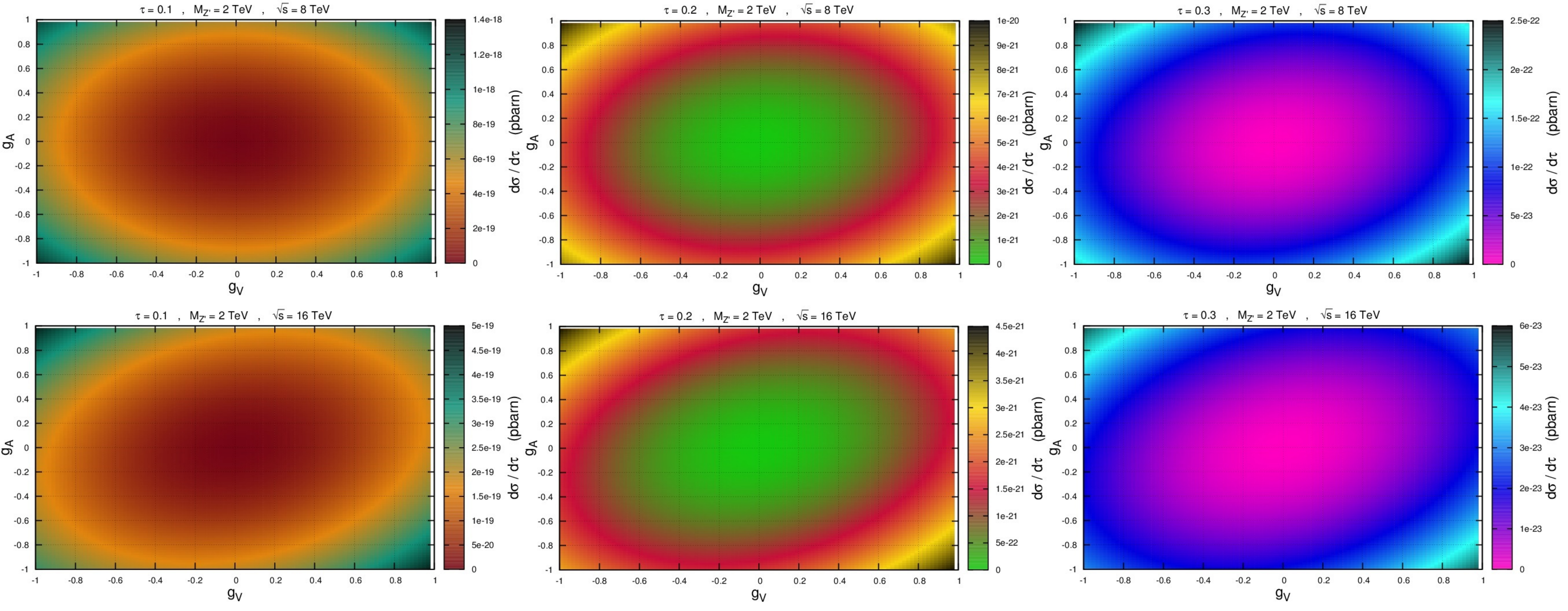}
\caption{\label{fig:gVgA-4}Process 2: Differential cross 
section with respect $\tau$ as a function of $g_V$ vs. $g_A$, for the 
$e-\mu$ decay mode of the $Z'$. 
We take $M_{Z'}$=2 TeV for $\tau$ = 0.1 (left), 0.2 (center), and 0.3 (right); 
the first row correspond to $\sqrt{s} = 8$, while the second row for 16 TeV.}
\end{figure}

When the branching ratio of the Higgs decaying into a pair of photons is
factored to the differential cross sections, it is assumed that there is no 
contribution arising from a new
$W'$ nor from new exotic fermions in the loop-level process $H \to \gamma \gamma$.
This can be show to be a pausible consideration by a seasoned choice of
quantum numbers for the fermion or because the existence of only an extra $U(1)'$
at the TeV level. Of course there is the necessity of supression for the
contribution coming from the charged Higgs sector. Still it is possible to
construct an specific model in which that it is allowed and pausible. 
For the case of $M_{Z'}= 2.5-3$ TeV, and a low value of $g''$, the coupling 
coming from the extra
$SU(2)$ or $U(1)$, we can have a expectation value of the extra Higgs fields,
that which generate the $Z'$ mass, of the order of 5-6 TeV. In this
way we found that the contribution of the charged Higgs sector is suppressed
in the $H \to \gamma\gamma$ decay.
\begin{figure}[!Ht]
\centering
\includegraphics[angle=0,scale=0.245]{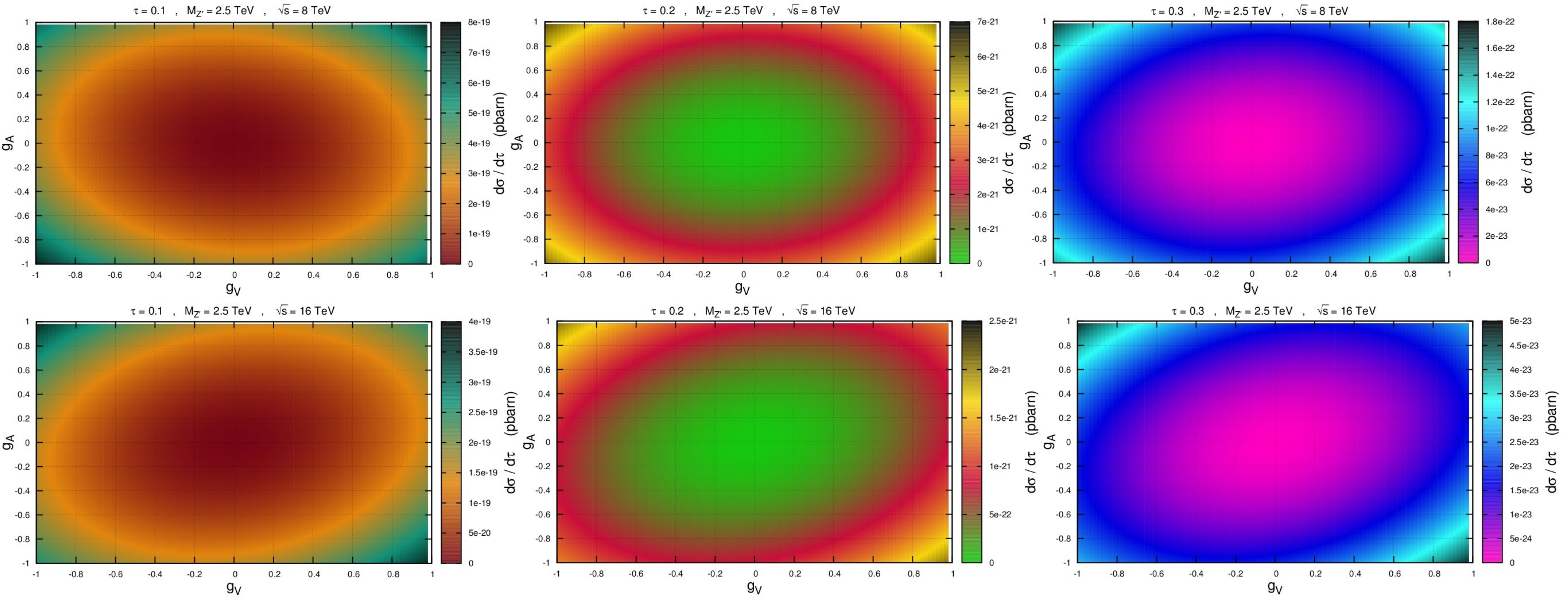}
\caption{\label{fig:gVgA-5}Process 2: Dependence of the differential cross 
section with respect to $g_V$ vs. $g_A$, for the $e-\mu$ decay mode of the $Z'$. 
We take $M_{Z'}$=2.5 TeV for $\tau$ = 0.1 (left), 0.2 (center), and 0.3 (right); 
the first row correspond to $\sqrt{s} = 8$, while the second row for 16 TeV.}
\end{figure}

\begin{figure}[!Ht]
\centering
\includegraphics[angle=0,scale=0.245]{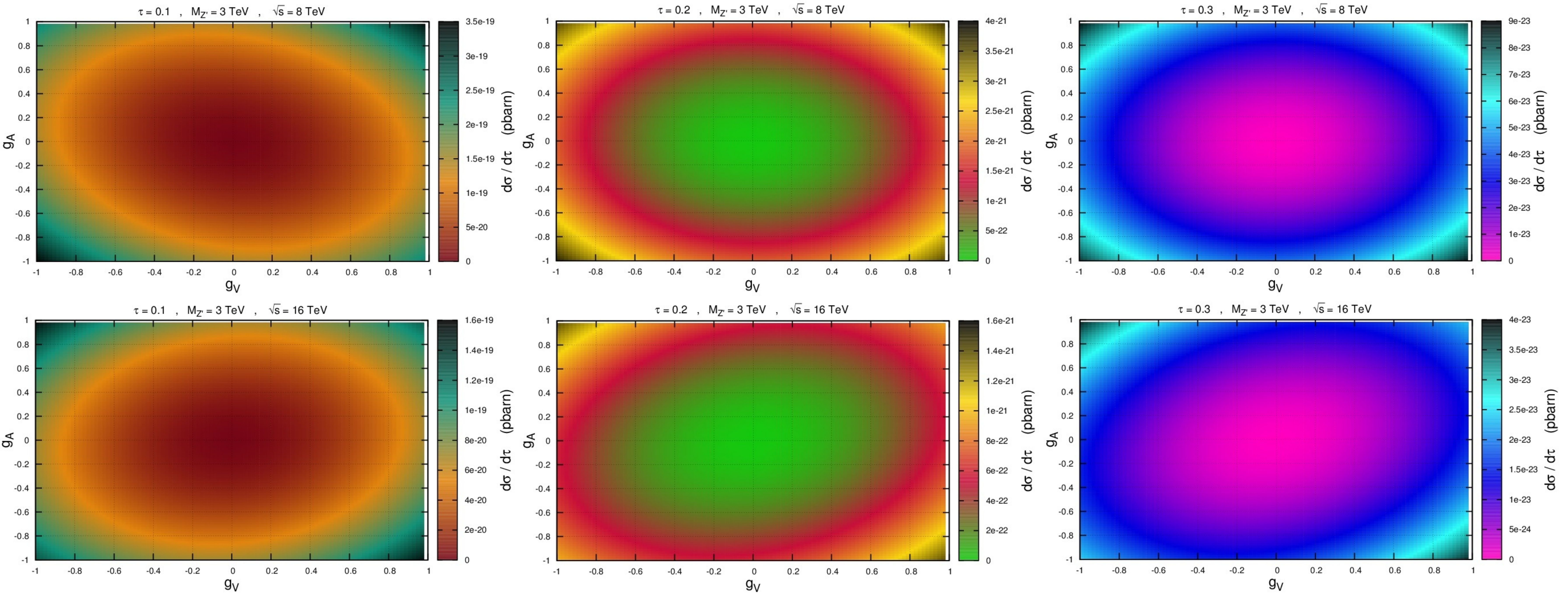}
\caption{\label{fig:gVgA-6}Process 2: Dependence of the differential cross 
section with respect to $g_V$ vs. $g_A$, for the $e-\mu$ decay mode of the $Z'$. 
We take $M_{Z'}$=3 TeV for $\tau$ = 0.1 (left), 0.2 (center), and 0.3 (right); 
the first row correspond to $\sqrt{s} = 8$, while the second row for 16 TeV.}
\end{figure}

Figs.~(11-13) show us the variation of the differential cross section, as a
function of $\tau$, on the values of the non-diagonal $g_V$ and $g_A$, in the
case of $e-\mu$ decay of $Z'$.

\section{Conclusions}
\label{sec:level4}

In this paper we have studied the possible signatures of the leptonic number 
violation vertex $\left(Z' l_{i}\bar{l}_{j}\right)$, as coming from $G(221)$ 
models with non-universal couplings \cite{Langacker:2000ju}, at LHC. The case 
$e \mu$ it is shown to be the most characteristic signal, since the cases
with a final $\tau$ could be more difficult to reconstruct from its decay 
products.
We have shown that the $e \mu$ signal could be feasible to be found at LHC, under 
reasonable expectations and conditions, in a foresable future. Our results could 
be shown to be of the same order of magnitude for similar processes studied in 
previous papers \cite{Jezo:2012rm}.

From the several energy distributions that we have shown, we can conclude that a 
possible leptonic number violation through a $Z'$, Drell-Yan-like, could be 
expected to be found in the case of a $Z'$ discovery. The events
differential distribution together with its distinctive leptonic signature make 
this kind of processes feasible to be found, if the 
nature has chosen this structure at the TeV scale. as we hope.

\acknowledgments We acknowledge support from CONACyT-SNI.

\bibliography{AllRef}
\bibliographystyle{JHEP}

\end{document}